\documentclass[aps,prr,superscriptaddress,twocolumn,longbibliography]{revtex4-2}

\usepackage[english]{babel}
\usepackage{graphicx}
\usepackage{amsmath}
\usepackage{xcolor}
\usepackage{mathtools}
\usepackage{amsfonts}
\usepackage{amssymb}
\usepackage{dsfont}
\usepackage{epstopdf}
\usepackage{subfigure}
\usepackage{pgf}
\usepackage{bm}
\usepackage{ulem}
\usepackage{verbatim}
\usepackage{hyperref} 

\newcommand{\ket}[1]{\ensuremath{\left|#1\right\rangle}}

\begin{document}
\title{Confinement and finite-range effects in a quasi-two-dimensional gas of fermionic dimers}
\date{June 17, 2026}

\author{Giovanni Midei} 
\affiliation{Departament de Fìsica, Universitat Politècnica de Catalunya, Campus Nord B4-B5, E-08034 Barcelona, Spain}
\affiliation{CQM group, School of Science and Technology, Physics Division, University of Camerino, Via Madonna delle Carceri, 9B, Camerino (MC), Italy}
\affiliation{INFN-Sezione di Perugia, 06123 Perugia, Italy}
\author{Jordi Boronat} 
\affiliation{Departament de Fìsica, Universitat Politècnica de Catalunya, Campus Nord B4-B5, E-08034 Barcelona, Spain}
\author{Grigory E. Astrakharchik}
\affiliation{Departament de Fìsica, Universitat Politècnica de Catalunya, Campus Nord B4-B5, E-08034 Barcelona, Spain}

\begin{abstract}
We investigate the ground-state properties of ultracold two-component Fermi gases in the presence of a transverse harmonic potential, focusing on the strongly interacting regime in which pairs of fermions form tightly bound molecules. Using the fixed-node diffusion Monte Carlo method, we calculate the equation of state and density profiles for the full fermionic system, which allows us to address the importance of finite-range corrections arising from the internal fermionic structure of the dimers. We interpret the results in terms of a molecular Bose gas in quasi-two-dimensional confinement and compare them with theoretical predictions for a weakly interacting two-dimensional Bose gas, identifying the range of validity of mean-field and beyond-mean-field descriptions. We also develop an analytical theory for the transverse density profile, capturing its broadening with increasing interaction strength. This work provides a benchmark for an effective bosonic description of strongly bound fermionic dimers and offers new insights into confinement-induced dimensional effects. 
\end{abstract}

\maketitle

In recent years, ultracold atoms have emerged as one of the most advanced experimental platforms for the precise exploration of quantum phenomena. In particular, experiments using magnetically controlled Feshbach resonances have allowed the efficient creation of dimers from strongly interacting atoms\cite{Kohler, Chin}. Deeply bound dimers can be treated as composite bosons, a feature which is particularly interesting when the constituent atoms are fermions\cite{Petrov} due to the change in quantum statistics and enhanced stability\cite{Regal, Strecker, Cubizolles, Jochim}. This has led to experimental observation of molecular Bose-Einstein condensation (BEC)\cite{Jochim1, Greiner, Zwierlein,Regal1} and superfluidity\cite{Zwierlein1}, as well as formulation of theories describing the crossover in strongly interacting Fermi gases from Bardeen–Cooper–Schrieffer (BCS) pairing to Bose–Einstein condensation (BEC)\cite{Zwerger, Strinati, Perali2004, Gaebler2010, Perali2011, Marsiglio2015} regimes. In these systems, collisional decay of molecules is strongly suppressed due to the Pauli exclusion principle\cite{Petrov}. Instead, when bosonic molecules are created in an atomic Bose gas\cite{Pilati2005}, this suppression is not present and molecules tend to have much shorter lifetimes\cite{Yurovsky}. Such collisional stability has recently been exploited to realize a stable repulsive Bose polaron in the strongly interacting regime\cite{Cesar}, hardly accessible in atomic Bose gases. Therefore, composite bosons made of paired fermions provide a robust platform for exploring the physics of strongly correlated bosonic systems. At the same time, residual effects arising from the composite nature of the molecular bosons might still play a role. In this context, quantum Monte Carlo methods are highly valuable for assessing the limits of applicability of mean-field theoretical descriptions. In fact, previous studies have examined in detail the molecular regime of ultracold fermions in both two\cite{Bertaina} and three\cite{Astra} dimensions. However, despite extensive studies of molecular Bose gases and their dimensional properties, the combined effects of the composite nature of fermionic dimers and the crossover from two to three dimensions remain largely unexplored. 

In the present Letter, we investigate a confined quasi-two-dimensional geometry, where effects arising from the underlying fermionic structure of the dimers and finite-range corrections become important. 
We use quantum Monte Carlo methods to calculate the ground-state energy and verify under which conditions and approximations it is possible to recover the equation of state of a two-dimensional Bose gas (2D). In order to describe how the transition from 2D to quasi-2D occurs, we calculate the density profile in the transverse direction. We show that occupation of the excited states of the harmonic confinement results in a density profile that approximately remains of a Gaussian shape but with a larger width. To this end, we develop an analytical description of this effect by minimizing the Gross–Pitaevskii (GP) energy functional and provide explicit expressions for the width and energy.

\begin{figure}[!t]
\includegraphics[width=\columnwidth]{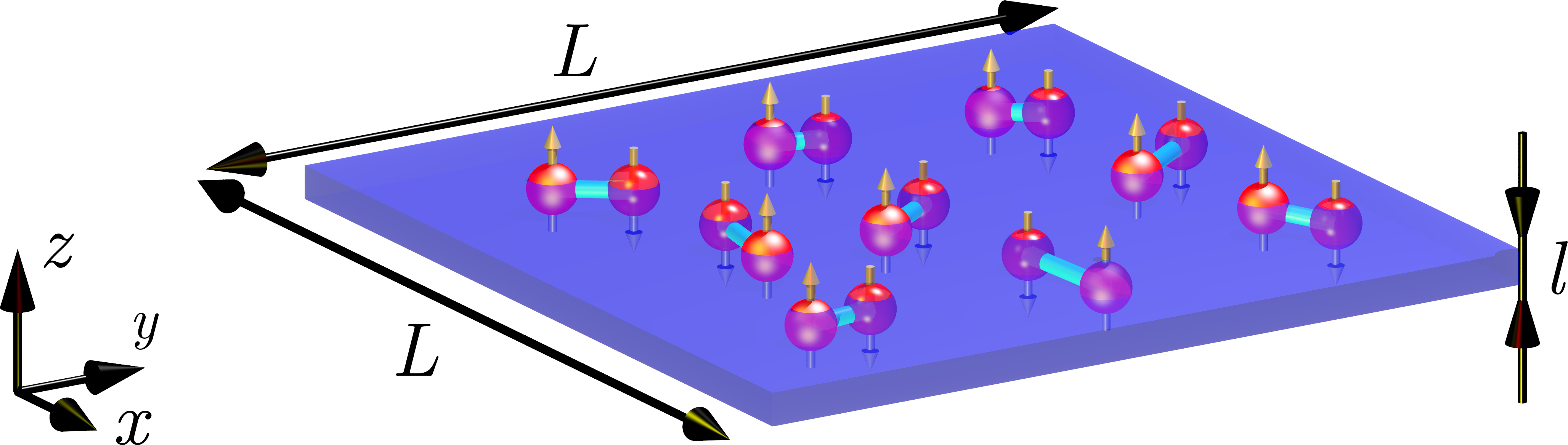}
\caption{Sketch of the system. Fermionic spin-up and spin-down atoms, represented as spheres with an arrow, are confined to a two-dimensional $(x,y)$ plane indicated by the blue rectangle. The finite width of the blue rectangle illustrates the harmonic confinement along the $z$ direction. Pairs of spin-up and spin-down fermions form tightly bound dimers, which can be approximately treated as composite bosons when dimer size $a_F$ is small compared to the mean interparticle distance, $n_F^{-1/2}$.}
\label{sketch}
\end{figure}

\begin{figure*}[!t]
\includegraphics[width=\textwidth]{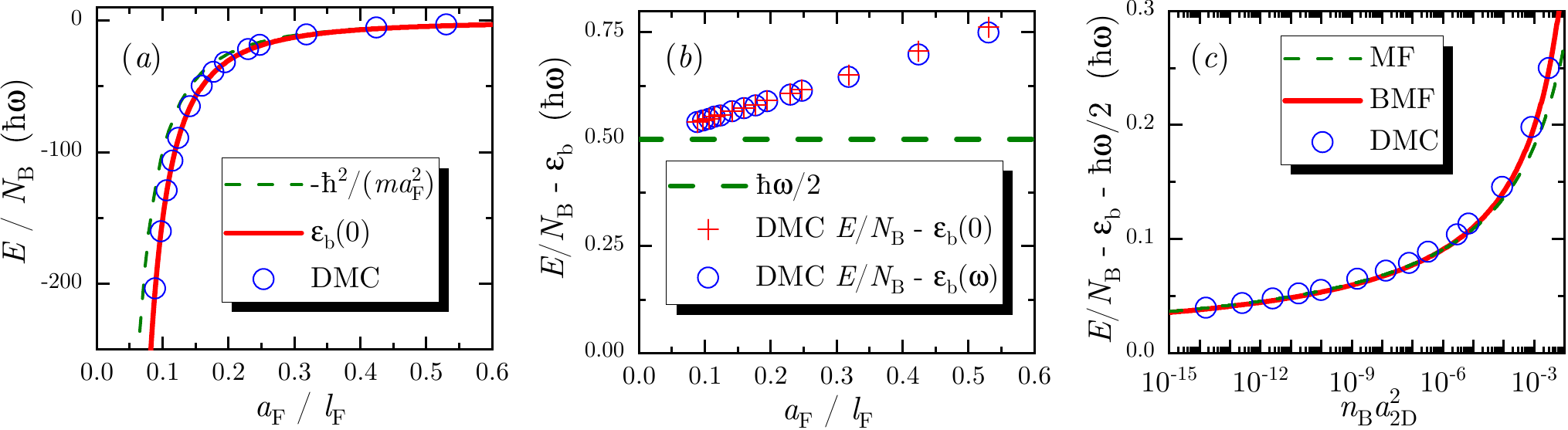}
\caption{Ground-state energy in units of harmonic oscillator level spacing. Panel a) energy per dimer obtained from FN-DMC calculation (blue circles), compared to the binding energy of two atoms interacting via square-well (red solid line) or contact (green dashed line) potentials as a function of the ratio between the 3D fermionic scattering length $a_F$ and oscillator length $l_F$.
Panel b) Energy after subtraction of the two–body binding energy in free space $\varepsilon_b(0)$ (red crosses) or in trapped geometry $\varepsilon_b(\omega)$ (blue circles). Ground–state energy of the harmonic oscillator $\hbar\omega/2$ is shown with a green dashed line.
Panel c) Energy of 2D motion as a function of the dimensionless 2D bosonic gas parameter $n_B a_{2D}^2$, compared with the mean–field (MF) (\ref{Eq:EoS:2D:MF}) and beyond–mean–field (BMF) (\ref{Eq:EoS:2D:BMF}) theoretical predictions for the 2D Bose gas energy. 
The FN DMC calculations are done using $N_F=66$ fermions.
}
\label{Fig2abc}
\end{figure*}

We consider a two-component non-polarized gas consisting of $N_{\uparrow}=N_{\downarrow}=N_F/2$ fermions of mass $m_F$, interacting via an attractive potential $V(r)$, in a transverse harmonic trap of frequency $\omega$, described by the following model Hamiltonian,
\begin{equation*}
\hat{H}\!=\!-\frac{\hbar^2}{2m_F}\!\!\left(\!\sum_{i=1}^{N_{\uparrow}}\!\nabla_i^2+\sum_{j=1}^{N_{\downarrow}}\!\nabla_j^2\!\!\right) + \sum_{i,j}\!V(\mathbf{r}_{ij})+\frac{1}{2}m_F\omega^2\!\sum_{i=1}^{N}\!\!z_i^2.
\end{equation*}
The gas, with density $n_F$, is confined by a harmonic potential along the $z$ direction, and periodic boundary conditions are imposed in the $(x,y)$ plane on a square box of area $L^2 = N_F/n_F$ (see the system sketch in Fig.~\ref{sketch}).
We model interspecies interactions using an attractive square-well potential, $V(r)=-V_0$, for $r < R_0$, and zero otherwise. 
The potential range $R_0$ is chosen to be small compared to the mean interparticle distance in the plane, $n_FR^2_0 = 10^{-3}$.
The $s$-wave fermionic scattering length $a_F$ is obtained by adjusting the depth of the square well $V_0$, according to 
$a_F = R_0 - \tan(K_0 R_0)/K_0$, where $K_0 = \sqrt{m_F V_0}/\hbar$ is the characteristic momentum associated with the potential.
For $K_0 R_0 > \pi / 2$, the scattering length is positive, $a_F > 0$, and a molecular state appears with its binding energy $\varepsilon^{(0)}_b$ determined by the transcendental equation $\sqrt{|\varepsilon^{(0)}_b|m_F}/\hbar\;R_0 \tan(K R_0)/(KR_0) = 1$,
with momentum $K = \sqrt{K_0^2 -|\varepsilon^{(0)}_b|m_F / \hbar^2}$.
However, due to the presence of the transverse harmonic confinement, the two–body binding energy is modified. 
For deeply-bound molecules, $\varepsilon^{(0)}_b \gg \hbar\omega$, the trap acts as a perturbation that shifts the bound–state energy by $\delta E (\omega)= \langle \psi^{(0)}_b |\frac{1}{2}\mu\omega^2 z^2| \psi^{(0)}_b \rangle / \langle \psi^{(0)}_b |\psi^{(0)}_b \rangle$, where $\mu=m_F/2$ is the reduced fermionic mass.
The unperturbed bound-state solution $\psi^{(0)}_b$ of the two-body Schrödinger equation with the square-well potential $V(r)$ is given by
\begin{equation}
\psi_b^{(0)}(r) =
\begin{cases}
  A \frac{\sin(Kr)}{r} & \text{if } r < R_0 \\
 \frac{e^{-\chi r}}{r}   & \text{otherwise}
\end{cases}
\end{equation}
where the constant $A=\frac{e^{-\chi R_0}}{\sin(KR_0)}$ is obtained by imposing the continuity of the wavefunction at $r=R_0$ and $\chi=\frac{\sqrt{m_F|\varepsilon^{(0)}_b|}}{\hbar}$.
Thus, the binding energy in a confined geometry is given by $\varepsilon_b (\omega) \approx \varepsilon_b^{(0)} + \delta E (\omega)$. 

We carried out simulations using the fixed-node diffusion Monte Carlo (FN-DMC) method\cite{Anderson1975, Anderson1980}. This technique yields the lowest energy compatible with the nodal surface of the many-body wave function. If the nodal surface ansatz is exact, the fixed-node energy would also be exact, otherwise the FN method provides an upper bound to the ground-state energy\cite{Reynolds1982}. We chose the guiding wave function as a product of one-body terms and a determinant of pairs,
\begin{equation}
\Psi_0({\bf r}_1,\cdots,{\bf r}_N) = \prod_{i=1}^{N} f_1(\mathbf{r}_i)\times\mathcal{A}\Big[\prod_{i<j}\psi^{(0)}_b(|\mathbf{r}_i-\mathbf{r}_j|)\Big] 
\label{eq2}
\end{equation}
where $\mathcal{A}$ denotes antisymmetrization, which ensures the Fermi-Dirac statistics under particle exchange. The one-body term, $f_1(\mathbf{r}_i)=\exp{(-\alpha z^2/l_F^2)}$, accounts for the external harmonic confinement along the $z$-direction, and the variational parameter $\alpha$ is optimized by minimizing the energy. For weak interactions, $\alpha \to 1/2$, corresponding to the single-particle ground state of a harmonic oscillator. We determined residual size effects by carrying out simulations with an increasing number of particles $N$ = 14, 38, and 66. We have verified that the finite-size correction to the energy for the considered particle numbers is below the reported statistical error. 

Deep into the BEC regime, tightly bound dimers behave as composite bosons with mass $m_B = 2 m_F$, density $n_B = n_F / 2$, and particle number $N_B = N_F / 2$. 
If the dimer size is small compared to the oscillator length $l_F$, the interactions between dimers remain governed by three-dimensional scattering. This allows one to relate the scattering length $a_B$ of bosonic dimers to the scattering length $a_F$ and the effective range $r_F$ of the fermionic atoms (for definitions of the scattering length and effective range, see, for example, Eq.~(8.76) in 3D and Eq.~(8.82) in 2D in Ref.~\cite{CastinBookTomeII}).  
Explicit relations between the atomic parameters ($a_F$ and $r_F$), the dimer parameters ($a_B$ and $r_B$), and the interaction potential details (interaction range $R_0$ and the momentum $K_0$ associated with the potential depth) are given in Refs.\cite{Kirk, Deltuva, Braaten01, Pera}
\begin{subequations}
\label{eq:11}
\begin{eqnarray}
r_F& =& R_0- \frac{R_0^3}{3a_F^2} - \frac{1}{K_0^2a_F}\label{eq:11a}\\
\frac{a_B}{a_F} &=& 0.5986(5) + 0.105\,\frac{r_F}{a_F}, \label{eq:11b} \\  
\frac{r_B}{a_F} &=& 0.133 + 0.51\,\frac{r_F}{a_F}, \label{eq:11c}
\end{eqnarray}
\end{subequations}
It can be anticipated that the composite nature of the molecular bosons manifests in finite-range corrections.
In the situation in which the transverse motion is frozen in the ground state, the system can be effectively described as being two-dimensional.
The relation between the 3D and 2D bosonic scattering lengths, including 3D finite–range corrections, is given by
\begin{equation}
a_{\mathrm{2D}} = l_{B}\, 2 e^{-\gamma}\sqrt{\frac{\pi}{\tilde A}}\, e^{-\sqrt{\frac{\pi}{2}}\, \frac{l_B}{a_B}},
\label{eq:a2D}
\end{equation}
where the effective range enters through the modified prefactor $\tilde A = A \exp\!\left(\sqrt{\pi/2}\, r_B/l_B\right)$ with $A = 0.905...$ corresponding to the zero effective range limit.

The system in the BEC limit possesses a double separation of scales, $\varepsilon_b \gg \hbar\omega \gg E_{2D}/N$, between the binding energy, harmonic oscillator level spacing, and the energy of the 2D motion. This allows one to develop an analytical perturbative description of the system, but also demands extremely precise numerical calculations.

In Fig.~\ref{Fig2abc}, we show FN-DMC results for the ground-state energy. The total energy is large and negative, reflecting the formation of dimers (see Fig.~\ref{Fig2abc}a). In the BEC regime, by subtracting the dimer energy, one gets a positive contribution which approaches the ground state energy of a single particle in a harmonic oscillator, $\hbar\omega/2$ (see Fig.~\ref{Fig2abc}b). Notice that for a precise comparison, it is important to consider the energy of dimers not in free space, but rather in the presence of transverse harmonic confinement (circles vs pluses in Fig.~\ref{Fig2abc}b).

Subtraction of $\hbar\omega/2$ finally allows us to obtain the energy associated with 2D motion, which can be conveniently presented as a function of the 2D gas parameter. 
\begin{figure}[h!]
\includegraphics[width=\columnwidth]{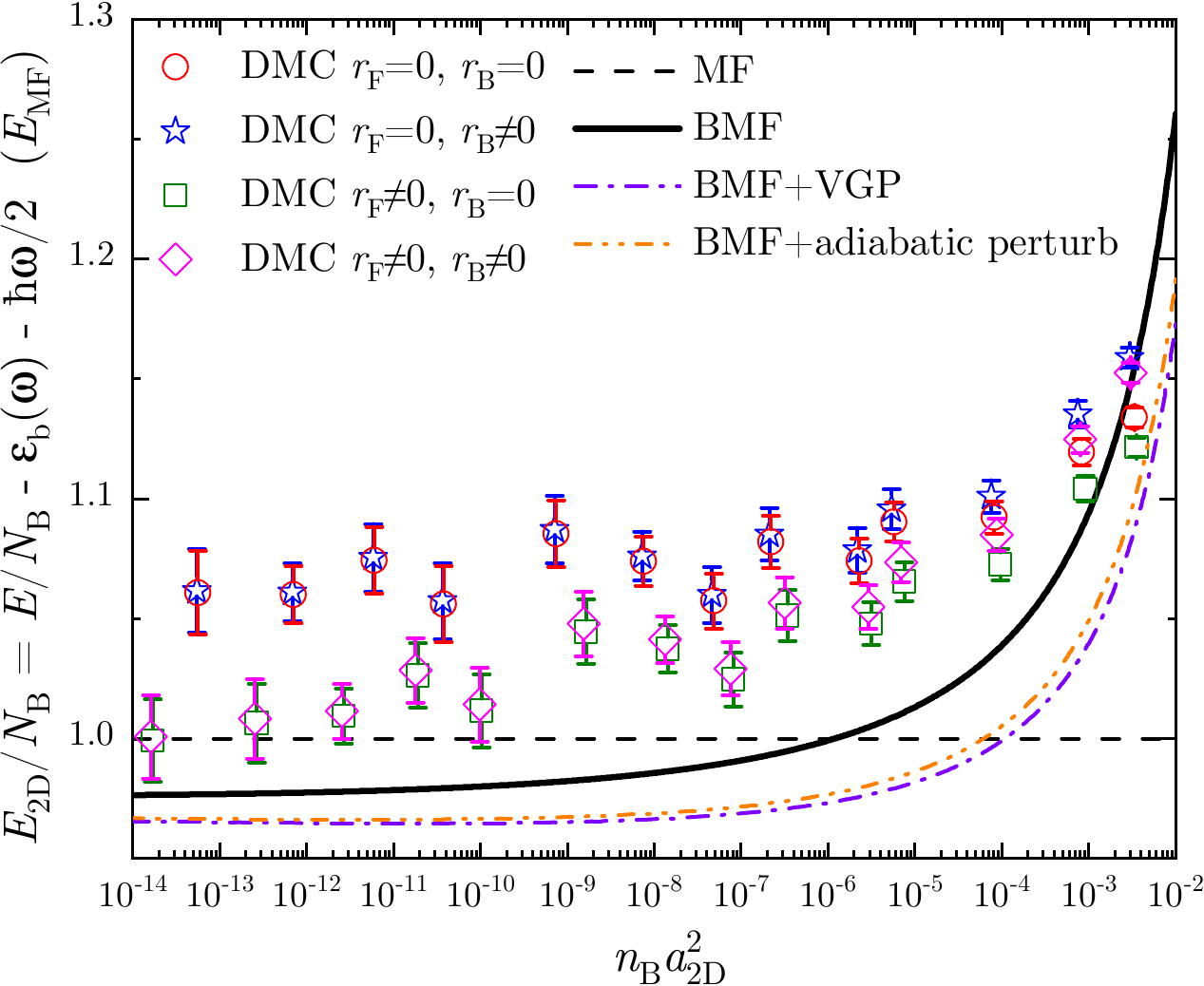}
\caption{Energy of the 2D motion per boson, $E_{2D}/N$, after subtracting both the two–body binding energy, $\varepsilon_b(\omega)$, and the harmonic–oscillator ground-state energy, expressed in units of the mean–field energy~(\ref{Eq:EoS:2D:MF}), $E_{MF}$. 
Symbols show DMC results obtained using different approximations for extracting $a_{2D}$ from Eqs.~(\ref{eq:11}).
Red circles -- both bosonic and fermionic ranges set to zero; 
blue stars -- finite bosonic range and zero fermionic range; 
green squares -- finite fermionic range and zero bosonic range; 
pink diamonds -- both ranges are finite. 
Analytic curves show different theoretical descriptions.
For a strictly 2D Bose gas, the mean-field (MF) Eq.~(\ref{Eq:EoS:2D:MF}), (dashed black line) and beyond-mean-field (BMF) Eq.~(\ref{Eq:EoS:2D:BMF}), (solid black line) contributions are shown for reference.
Quasi-2D corrections are included on top of the BMF theory by subtracting the negative contribution obtained either from variational Gross–Pitaevskii (VGP) theory [Eq.~(\ref{Eq:Q2Denergy}), dash-dotted line] or from adiabatic perturbation theory [Eq.~(\ref{olsh}), dash-dot-dotted line].
}
\label{2}
\label{Fig3}
\end{figure}
In the dilute regime, we find good agreement with the mean-field energy \cite{Astra1, Pilati2005}, 
\begin{eqnarray}
\frac{E_{MF}}{N_B} = \frac{2\pi\hbar^2n_B/m_B}{|\ln n_Ba_{2D}^2|}
\label{Eq:EoS:2D:MF}
\end{eqnarray} 
which exhibits a weak (logarithmic) dependence on the scattering length. 
At the same time, the 2D scattering length has an exponential dependence on the 3D fermionic scattering length (\ref{eq:a2D}) (compare horizontal axis in Fig.~\ref{Fig2abc}b and Fig.~\ref{Fig2abc}c, which show the same data points). 
For large values of the gas parameter, we observe beyond-mean-field (BMF) corrections 
\begin{eqnarray}
\!\frac{E_{BMF}}{N_B}\!=\! 
\frac{2\pi\hbar^2 n_B/m_B}{
\!|\!\ln\!n_{\!B}a_{\!2D}^2|\!+\!\ln\!|\!\ln n_{\!B}a_{\!2D}^2\!|\!+\!C^E_1\!+\!\frac{\!\ln\!|\!\ln\!n_{\!B} a_{\!2D}^2\!|\!+\!C^E_2\!}{
|\ln n_B a_{2D}^2|
} 
}
\label{Eq:EoS:2D:BMF}
\end{eqnarray}
with coefficients
$C^E_1 =  -\ln\pi -2\gamma -1/2 = - 2.80$ and\\
$C^E_2 =  -\ln\pi -2\gamma +2.0(1) +1/4= -0.05(10)$\cite{Astra1}.  

To examine the importance of the corrections arising from finite fermionic and bosonic effective ranges, it is convenient to normalize the energy of 2D motion by the mean-field value, thereby magnifying the difference.
\begin{figure}[h!]
\includegraphics[width=\columnwidth]{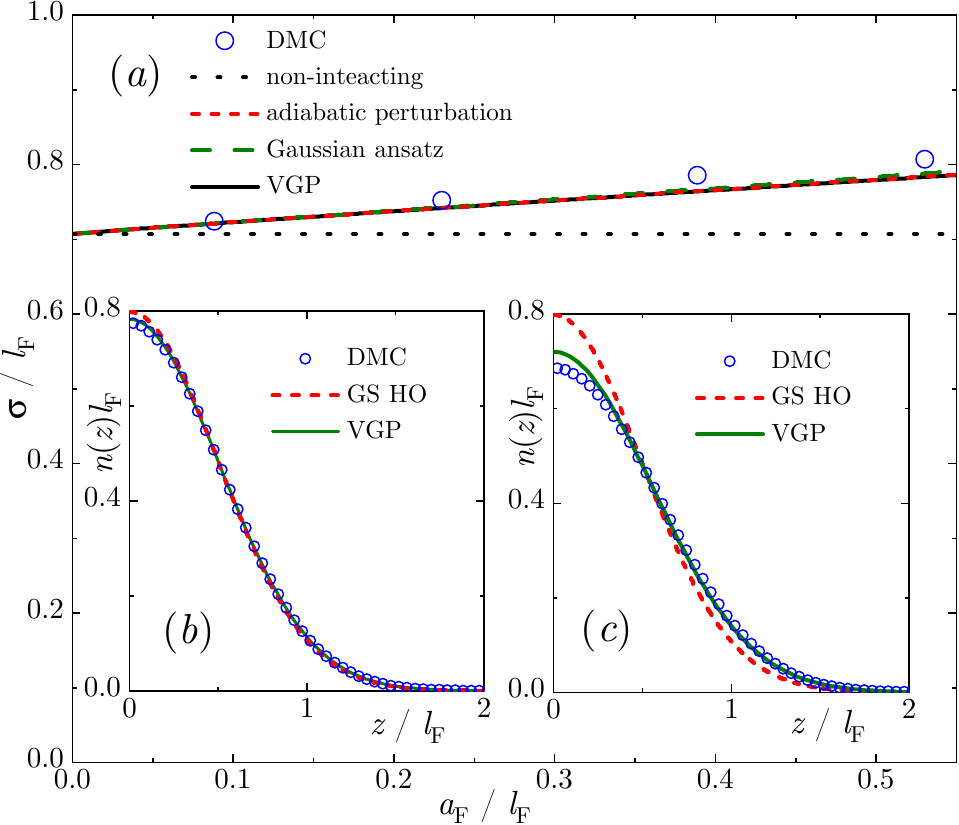}
\caption{
Transverse density profile and transverse broadening. 
Panel (a): Width $\sigma$ of the Gaussian density profile, in units of the fermionic oscillator length $l_{\mathrm{F}}$, as a function of the 3D fermionic scattering length $a_F$ in units of $l_{\mathrm{F}}$. The black dotted line corresponds to the case of constant $\sigma = l_B$, the solid black line shows the numerical solution obtained from the minimization of the Gross–Pitaevskii functional, the dotted green line is the analytical solution in Eq.~\eqref{GP} and the dashed red line is the solution obtained from adiabatic perturbation theory ~\cite{Merloti}, that coincides with VGP to lowest order, as shown in Appendix~\ref{C}. Blue circles represent $\sigma$ values extracted from a Gaussian fit to the DMC density profile along the $z$ direction. The insets display the DMC density profiles along $z$ at $a_F/l_{F}=0.09$ (b) and $a_F/l_{F}=0.53$ (c), compared with the variational Gross–Pitaevskii (VGP) profiles and with the harmonic–oscillator ground–state density.
}
\label{3}
\end{figure}
The resulting comparison is presented in Fig.~\ref{2}.
We find that in the dilute regime ($n_Ba_{2D}^2\lesssim 10^{-6}$), it is important to take into account the fermionic effective range, as it provides a substantial correction that reduces the energy and makes it approach the BMF equation of state.
This reflects that dimers are far apart and the energy is primarily sensitive to the internal structure of each composite boson. 
However, as the gas parameter increases ($n_Ba_{2D}^2\gtrsim 10^{-6}$), dimers come closer, and the bosonic range correction becomes relevant and must be included. 

Another key phenomenon determining the system properties is the modification of the transverse structure induced by tight confinement. 
To quantify this effect, we calculate the density profile $n(z)$ along the transverse direction.
We extract the effective width $\sigma$ by fitting Monte Carlo data and report it in Fig.~\ref{3}a. Deeply in the 2D regime, only the lowest state of the transverse harmonic oscillator is occupied, resulting in a Gaussian shape, $n(z) \propto \exp(-z^2/l)$, with its width fixed by the oscillator length $l = \sqrt{\hbar/m\omega}$, see Fig.~\ref{3}b. 
Since the dimer mass is twice the atomic one, the density profile is narrower in the BEC limit than in the unitary regime, where the width is $l=l_B = l_F / \sqrt{2} $. With increasing interactions between dimers, the 2D motion energy eventually becomes comparable to the harmonic oscillator level spacing $\hbar\omega$, and atoms start to populate higher states of the harmonic oscillator. 
A typical example of the density profile in that regime is shown in Fig.~\ref{3}c. 
We find that for a weak population of transverse states, the density profile is still well approximated by a Gaussian, but with a wider width $\sigma>l_B$.
In addition, we develop and test several analytical approaches based on the Gross–Pitaevskii (GP) energy functional to describe the transition between the 2D and quasi-two-dimensional regimes.
In particular, we find that the following approaches have comparable accuracy:
(i) minimizing the GP energy functional assuming a Gaussian ansatz, resulting in
\begin{equation}
\sigma = l_B + \sqrt{\frac{\pi}{2}}\,n_{B}\,a_B\,l_B^{2}
\label{GP}
\end{equation}
This expression is valid for $n_{B}\,a_B\,l_B \ll 1$, where the width remains close to the harmonic oscillator length $l_B$. The derivation of Eq.~(\ref{GP}) is given in Appendix~\ref{A}.
Deviations from the exact numerical solution remain below $1\%$ for the range of scattering lengths $a_F$ considered in Fig.~\ref{3}, confirming the accuracy of the perturbative expansion in this regime. The broadening of the density profile reflects the onset of three-dimensional behavior, as the molecular gas begins to explore higher transverse modes beyond the strictly 2D regime. 
(ii) Expanding the wavefunction in harmonic oscillator modes as in Ref.~\cite{Merloti} and using their approach to calculate the variance in $z$ direction, $\langle z^2\rangle =\sigma^2/2$. In this case we derive the width $\sigma$ from the perturbative ground-state solution of the one-dimensional Gross–Pitaevskii equation along the confined direction, within the framework of adiabatic perturbation theory using the Hellmann--Feynman theorem (for details, see Appendices~\ref{B} and \ref{C}). We find that to leading order it coincides with the result from the variational Gaussian ansatz.
This agreement is notable because the variational approach assumes a Gaussian profile, whereas the perturbative theory accounts for all harmonic oscillator levels. 
From the energetic point of view, due to the residual repulsive interaction between the bosons, occupation of additional degrees of freedom lowers the system energy,  see dash-dot and dash-dot-dot lines in Fig.~\ref{Fig3}. 
For the considered parameters, essentially only $n=2$ state is excited while occupation of $n=4$ and higher states is negligible (see  Appendix~\ref{D} for the calculation of the projection coefficients).

In conclusion, we perform FN-DMC calculations for the ground state of a two-component Fermi gas in the BEC regime under tight harmonic confinement and discuss finite-range corrections arising from the internal fermionic structure of the composite bosons. Remarkably, the results show excellent agreement with mean-field and beyond-mean-field predictions, despite the total energy being about two orders of magnitude larger than the energy of the two-dimensional Bose gas, obtained by subtracting the molecular and trap contributions. 
In addition, we study how the system exits the two-dimensional regime and starts to occupy the transverse direction.  We develop an analytical theory for the transverse density profile using the three-dimensional Gross–Pitaevskii energy functional, and find a close agreement with FN-DMC results. Future work might include the investigation of the system for stronger interactions approaching the unitary limit, and the exploration of the BCS regime. This work may provide new insights into bosonic molecular systems composed of fermions, which are of particular interest due to their enhanced stability compared to atomic Bose gases.

{\it Acknowledgements ---} We are grateful to Y. Castin, J. Levinsen, A. Perali, D.S. Petrov, and S. Pilati for fruitful discussions. G.M. would like to thank the Barcelona Quantum Monte Carlo (BQMC) group for warm hospitality and support during a six-month visit to the Universitat Politècnica de Catalunya (UPC), Barcelona, Spain. G.M. also acknowledges financial support from the INFN and the University of Camerino. We also acknowledge access to supercomputer resources as provided through grants from the Red Española de Supercomputación (FI-2025-1-0020) and financial support from the Ministerio de Ciencia e Innovación MCIN/AEI/10.13039/501100011033 (Spain) under Grant No. PID2023-147469NB-C21.

%

\clearpage

\appendix

\section{Gross-Pitaevskii energy functional}
\label{A}
We start from the 3D Gross-Pitaevskii energy functional for a Bose gas with contact interactions confined along $z$,
\begin{equation}
E\!=\!\!\int\!\!d^3r\!\!\left[\frac{\hbar^2|\nabla\psi(\mathbf{r})|^2}{2m_B}\!+\!\frac{1}{2} m_B\omega_z^2 z^2|\psi(\mathbf{r})|^2\!+\!\frac{g_{3D}}{2} |\psi(\mathbf{r})|^4 \right],
\label{VGPeq}
\end{equation}
where $g_{3D} = 4 \pi \hbar^2 a_B/m_B$ is the coupling constant. The condensate wavefunction is factorized, $\psi(\mathbf{r}) = \sqrt{n_{B}} \, \phi(z)$, where $n_{B}$ denotes the constant 2D bosonic density in the $xy$-plane while for $\phi(z)$ we assume a Gaussian shape of width $\sigma$,
\begin{equation}
\phi(z) =  \frac{1}{(\pi \sigma^2)^{1/4}}\exp\left(-\frac{z^2}{2\sigma^2}\right)
\label{wf}
\end{equation}
The resulting total energy per number of bosons
\begin{equation}
\frac{E(\sigma)}{N_{B}} =  \frac{\hbar^2}{4m_B} \frac{1}{\sigma^2} + \frac{1}{4} m_B \omega^2 \sigma^2 + \frac{g_{3D}}{2 \sqrt{2 \pi} \sigma} n_{B}.
\label{Eq:VGP:E}
\end{equation}
depends explicitly on the variational width $\sigma$ of the Gaussian along $z$. 
Minimization of the total energy with respect to $\sigma$ gives the following equation
\begin{equation}
m_B \omega_z^2 \sigma^4 - \frac{g_{3D}}{ \sqrt{2 \pi}} n_{B} \sigma - \frac{\hbar^2}{m_B} = 0.
\label{eq:quartic}
\end{equation}
The only real and positive solution of this equation for $\sigma$ determines the optimal width of the Gaussian density profile along $z$. For vanishing interaction $g_{3D}=0$, the solution to Eq.~\eqref{eq:quartic} corresponds to the bosonic oscillator length $\sigma_0 = \sqrt{\hbar/(m_B\omega_z)} = l_B$.
We now include weak interactions by writing $\sigma = \sigma_0 + \delta$, with $ \delta \ll \sigma_0$. Substituting this into Eq.~\eqref{eq:quartic} and expanding in powers of $\delta$ we obtain an analytical expression for the variational width that minimizes the total energy
\begin{equation}
\sigma \simeq {l_B} + \sqrt{\frac{\pi}{2}}\,n_{B}\,a_B\,l_B^2.
\label{VGPP}
\end{equation}
which is valid if the condition $\delta\ll\sigma_0$ is satisfied, i.e, $n_{B}\,a_B\,l_B\ll1$. 

Using the perturbative expansion~(\ref{VGPP}), we explicitly express the negative energy correction in the GP functional arising from the occupation of the transverse degrees of freedom,
\begin{equation}
\frac{E}{N_B} = \frac{\hbar\omega}{2} + \frac{g_{2D}}{2}n_{B} - \frac{\pi}{2} \frac{ \hbar^2a_B^2}{m_B} \,n_{B}^{2},
\label{Eq:Q2Denergy}
\end{equation}
where $g_{2D}=2\sqrt{2\pi}\hbar^2 a_B/(m_B l_{B})$ is the two-dimensional coupling constant.

\section{Adiabatic perturbation theory}
\label{B}
In this Section, we compare the estimation of the ground state energy of the Bose gas with the results obtained in Ref.~\cite{Merloti}, where the first-order correction to the chemical potential of the gas due to the presence of the harmonic trap has been added on top of the unperturbed mean field solution, yielding
\begin{equation}
\mu(n_{B})= \frac{\hbar\omega}{2} + g_{2D}\,n_{B} - \frac{16\pi^{2}\hbar^{2}a_B^{2}|c_2|}{m_B}\,n_{B}^{2},
\label{eqmerloti}
\vspace{0.3cm}
\end{equation}
where $c_2=-0.033$ is a constant obtained from the summation over excited states of the harmonic oscillator. The energy 
can be obtained by integration of the chemical potential, 
$\varepsilon(N) = \int_{0}^{N}\mu(n')\, dN'$. The resulting energy per particle is
\begin{equation}
\frac{E}{N_B} = \frac{\hbar\omega}{2} + \frac{g_{2D}}{2}n_{B} - \frac{16\pi^2}{3} \frac{\hbar^2a_B^2|c_2|}{m_B}n_{B}^{2} .
\label{olsh}
\end{equation}

\section{Hellmann--Feynman theorem and trap energy}
\label{C}
We now consider the application of the Hellmann--Feynman (HF) theorem to separate the different energy contributions in the trapped system. 
Let $\hat H(\lambda)\ket{\psi(\lambda)} = E(\lambda)\ket{\psi(\lambda)}$, where $\hat H(\lambda)$ is a Hermitian operator and $\ket{\psi(\lambda)}$ a non-degenerate eigenstate of that operator. 
The HF theorem states
\begin{equation}
\frac{dE}{d\lambda} = \Big\langle \psi(\lambda)\Big|\frac{\partial \hat H}{\partial \lambda}\Big|\psi(\lambda)\Big\rangle.
\end{equation}
In our case, the Hamiltonian reads
$\hat H = \hat T + \hat V_{\mathrm{trap}} + g_{\mathrm{3D}}\hat V_{\mathrm{int}},$
so that the potential energy contribution due to the harmonic trap is given by $E_{\mathrm{trap}} = \langle \hat V_{\mathrm{trap}}\rangle=m_B \omega^2  \langle \hat{z}^2 \rangle /2$. Applying the HF theorem, one obtains the relation which connects the potential energy due to the trap to the derivative of the total energy in Eqs.~\eqref{olsh} with respect to the trapping frequency
\begin{equation}
E_{\mathrm{trap}}
= \frac{\omega}{2}\frac{\partial E}{\partial \omega}= \frac{\hbar\omega}{4}
+ \frac{n_{B}\,g_{\mathrm{3D}}\sqrt{m_B\omega}}{8\sqrt{2\pi\hbar }}.
\label{eq:Etrap_def}
\end{equation}
Since $\langle \hat{z}^2 \rangle=\sigma^2/2$, where the bracket is taken between the Gaussian variational states in Eq.~\eqref{wf}, we have 
\begin{equation}
\sigma^2= 2\langle z^2\rangle
= \frac{4E_{\mathrm{trap}}}{m_B\omega^2}
= \frac{\hbar}{m_B\omega}
+\frac{n_{B}\,g_{\mathrm{3D}}}{2\sqrt{2\pi\,\hbar m_B\,\omega^{3}}}.   
\end{equation}
Writing explicitly $g_{\mathrm{3D}}$ and
$\omega=\hbar/(m_B l_B^2)$, we obtain
\begin{equation}
\sigma
= l_B\sqrt{1\;+\;\sqrt{2\pi}\,a_
B\;n_{B}\,l_B}
\end{equation}
To leading order in the small parameter $a_B n_{2D}\,l_B \ll 1$, we recover Eq.~\eqref{VGPP} obtained using the VGP approach. This agreement is nontrivial, because the variational approach assumes a Gaussian profile, whereas the perturbative treatment includes all harmonic-oscillator levels.
 Therefore, both methods yield the same first-order correction to the width $\sigma$. By contrast, the leading correction to the energy is different, as can be seen by comparing Eq.~\eqref{Eq:Q2Denergy} with Eq.~\eqref{olsh}. The reason is that the coefficient $c_2$, which accounts for all harmonic-oscillator levels in adiabatic perturbation theory, enters only the interaction energy and does not affect the trap contribution from which $\sigma$ is determined. As a consequence, the two approaches give the same perturbative expression for $\sigma$, but different ones for the energy $E$.

\section{Occupation of excited transverse harmonic-oscillator levels}
\label{D}
Of particular interest is how the single-particle harmonic-oscillator states become populated as the system ceases to be purely 2D and begins to excite transverse levels of the harmonic oscillator. 
To quantify the population of the excited states we compute the overlap 
$c_n = \int_{-\infty}^{\infty} \phi(z)\,\varphi_n(z)\,dz$
between the broadened wave function~(\ref{wf}) and the $n$-th state of the harmonic oscillator,  
$\varphi_n(z) = \frac{1}{\sqrt{2^n n!\, a_{ho}}}\, e^{-z^2/(2a_{ho}^2)}\, H_n\!\left(\frac{z}{a_{ho}}\right), \quad n=0,1,\ldots,$
where $H_n(x)$ are Hermite polynomials and $a_{ho}$ is the oscillator length.
The overlap can be calculated explicitly
\begin{eqnarray}
c_n^2 =
\begin{cases}
\displaystyle
\frac{n!}{2^{\,n-1}\,\left(\frac{n}{2}!\right)^2}
\,\frac{a_{\mathrm{ho}}\sigma\,(\sigma^2 - a_{\mathrm{ho}}^2)^{n}}
{(\sigma^2 + a_{\mathrm{ho}}^2)^{n+1}},
& \text{even $n$}, \\[10pt]
0, & \text{odd $n$}.
\end{cases}
\label{c2}
\end{eqnarray}
and satisfies $\sum_{n=0}^\infty c_n^2=1$. 
We find that, for weak interactions, essentially only the $n=2$ state is occupied (with a typical occupation of a few percent), while higher states have small amplitudes over the considered range of parameters. 
For example, for the parameters of Fig.~\ref{3}c, Eqs.~(\ref{GP},\ref{c2}) predict $c_2^2=0.0069$ for the occupation of the $n=2$ state, which is close to the occupation obtained from the density profile $n(z)$ according to  
$c_n = \int_{-\infty}^{\infty} \sqrt{n(z)}\,\varphi_n(z)\,dz$
which results in $c_2^2 = 0.0073$. 
That is Eq.~(\ref{c2}) can be used to estimate the occupations of the excited states of the transverse harmonic oscillator. 
\end{document}